\documentclass[pra,twocolumn,10pt,showkeys,showpacs,amsmath,amsfonts,floatfix,superscriptaddress,nofootinbib,aps]{revtex4-1} 
\usepackage{amssymb,amsmath,amstext}                
\usepackage{graphicx}                                               
\usepackage{color}       
\usepackage{bm}
\usepackage{appendix}                                              
\usepackage[latin2]{inputenc}
\usepackage{ulem}
\usepackage{latexsym}
\usepackage{tabularx}
\usepackage{afterpage}
\usepackage{nonfloat}
\usepackage[colorlinks=true,citecolor=blue,linkcolor=magenta]{hyperref}
\usepackage{lipsum}

\def\be{\begin{equation}}
\def\ee{\end{equation}}
\def\bea{\begin{eqnarray}}
\def\eea{\end{eqnarray}}
\def\bi{\begin{itemize}}
\def\ei{\end{itemize}}

\newcommand{\ket}[1]{\mbox{$| #1 \rangle$}}

\begin{document}

\title{
Monte Carlo sampling from a projected entangled-pair state \\
in simulations of quantum annealing in the three dimensional random Ising model 
}

\newcommand{\affilju}{
             Jagiellonian University, 
             Faculty of Physics, Astronomy and Applied Computer Science,
             Institute of Theoretical Physics, 
             ul. \L{}ojasiewicza 11, 30-348 Krak\'ow, Poland 
             }

\newcommand{\affilkac}{  
             Jagiellonian University, 
             Mark Kac Center for Complex Systems Research,
             ul. \L{}ojasiewicza 11, 30-348 Krak\'ow, Poland 
             }
             
\author{Jacek Dziarmaga}\affiliation{\affilju}\affiliation{\affilkac}

\date{June 1, 2026}

\begin{abstract}
Quantum annealing with the D-Wave Advantage system in the random Ising model on a cubic lattice is simulated using a three-dimensional (3D) tensor network. The Hamiltonian is driven across a quantum phase transition from a paramagnetic phase to a spin-glass phase. The network is represented as a tensor product state, also known-particularly in two dimensions-as a projected entangled-pair state (PEPS).
The annealing procedure is repeated for a range of annealing times in order to test the Kibble-Zurek (KZ) power law governing the residual energy at the end of the annealing ramp. For an infinite lattice with periodic nearest-neighbor random Ising couplings, the final energy is evaluated using a deterministic method. For a finite lattice with open boundaries, we introduce a more efficient Monte Carlo sampling approach.
In both cases, the residual energy as a function of annealing time approaches the KZ power law as the annealing time increases.
\end{abstract}

\maketitle


\section{Introduction}
\label{sec:intro}

The Kibble-Zurek mechanism (KZM), originally proposed to describe the formation of topological defects during cosmological phase transitions~\cite{K-a,*K-b,*K-c}, was later adapted to continuous phase transitions in condensed matter systems. In this context, it evolved into a dynamical theory predicting how the density of defects scales with the quench rate according to the universality class of the transition~\cite{Z-a,*Z-b,*Z-c,Z-d}. These predictions have been confirmed in numerous numerical studies~\cite{KZnum-a,KZnum-b,KZnum-c,KZnum-d,KZnum-e,KZnum-f,KZnum-g,KZnum-h,KZnum-i,KZnum-j,KZnum-k,KZnum-l,KZnum-m,that,KZ_weakly_first} and experiments~\cite{KZexp-a,KZexp-b,KZexp-c,KZexp-d,KZexp-e,KZexp-f,KZexp-g,QKZexp-a,KZexp-h,KZexp-i,KZexp-j,KZexp-k,KZexp-l,KZexp-m,KZexp-n,KZexp-o,KZexp-p,KZexp-q,lamporesi2013,donadello2016,KZexp-s,KZexp-t,KZexp-u,KZexp-v,KZexp-w,KZexp-x}.
The quantum version of the mechanism (QKZM) describes quenches across quantum critical points in isolated systems~\cite{QKZ1,QKZ2,QKZ3,d2005,d2010-a,d2010-b,QKZteor-a,QKZteor-b,QKZteor-c,QKZteor-d,QKZteor-e,QKZteor-f,QKZteor-g,QKZteor-h,QKZteor-i,QKZteor-j,QKZteor-k,QKZteor-l,QKZteor-m,QKZteor-n,QKZteor-o,KZLR1,KZLR2,QKZteor-oo,delcampostatistics,KZLR3,QKZteor-q,QKZteor-r,QKZteor-s,QKZteor-t,sonic,QKZteor-u,QKZteor-v,QKZteor-w,QKZteor-x,roychowdhury2020dynamics,sonic,schmitt2021quantum,RadekNowak,dziarmaga_kinks_2022,transverse_oscillations,persistent_osc,delcampomagic}. Its predictions have also been tested experimentally~\cite{QKZexp-a,QKZexp-b,QKZexp-c,QKZexp-d,QKZexp-e,QKZexp-f,QKZexp-g,deMarco2,Lukin18,adolfodwave,2dkzdwave,King_Dwave1d_2022,Semeghini2021,Satzinger2021etal}.

The basic idea of QKZM can be summarized as follows (see also Refs.~\onlinecite{d2010-a,d2010-b,Z-d,ROSSINI20211} for reviews). A smooth ramp crossing the critical point at time $t_c$ can be linearized near the transition as
\begin{equation}
\epsilon(t)=\frac{t-t_c}{\tau_Q}.
\label{epsilont}
\end{equation}
Here $\epsilon$ is a dimensionless parameter in the Hamiltonian that measures the distance from the quantum critical point, and $\tau_Q$ is the KZ quench time. Initially, the system is prepared in its ground state and follows the instantaneous ground state adiabatically far from the transition. Adiabaticity breaks down near the time $t-t_c=-\hat t$, when the energy gap becomes comparable to the quench rate,
\begin{equation}
\Delta\propto|\epsilon|^{z\nu} \propto |\dot \epsilon/\epsilon| = 1/|t-t_c|.
\end{equation}
This condition defines the characteristic timescale
$
\hat t \propto \tau_Q^{z\nu/(1+z\nu)},
$
where $z$ and $\nu$ are the dynamical and correlation-length critical exponents, respectively. The corresponding KZ correlation length,
\begin{equation}
\hat\xi \propto \hat t^{1/z} \propto \tau_Q^{\nu/(1+z\nu)},
\label{eq:hatxi}
\end{equation}
defines the characteristic length scale of KZ excitations after crossing the critical point.

\begin{figure}[t!]
\includegraphics[width=0.8\columnwidth]{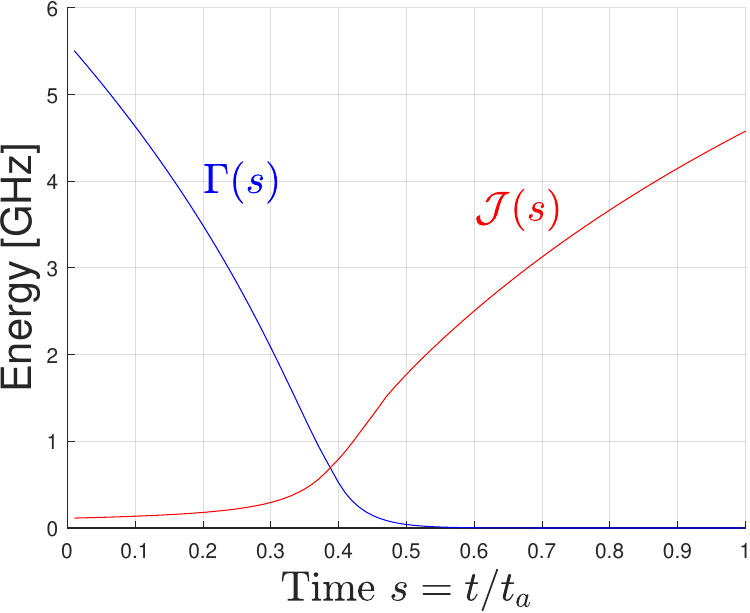}
\caption{
{\bf Quantum annealing ramp. } 
The energy scales $\Gamma$ and ${\cal J}$ (in GHz) in the Hamiltonian \eqref{eq:H} are shown as functions of the dimensionless ramp parameter $s=t/t_a$, where $t_a$ is the annealing time.
This corresponds to the annealing schedule of the D-Wave Advantage system used in Ref.~\onlinecite{Science_Dwave}.
}\label{fig:QA}
\end{figure}

Recent progress in programmable quantum simulators has opened new opportunities for exploring QKZM in higher dimensions. In particular, Rydberg-atom arrays enable versatile emulation of quantum many-body systems~\cite{rydberg2d1,rydberg2d2,Semeghini2021,Satzinger2021etal}, while coherent D-Wave devices provide another platform for studying nonequilibrium quantum dynamics~\cite{King_Dwave1d_2022,King_Dwave_glass,Science_Dwave}. These systems make it possible to investigate QKZM in two- and three-dimensional (3D) settings and to use its scaling predictions as a probe of the quantumness of the hardware~\cite{RadekNowak,King_Dwave1d_2022,dziarmaga_kinks_2022,schmitt2021quantum,transverse_oscillations,persistent_osc,dwave_mitigation,XX_Google,XX_Google_JU}.

Quantum simulations can be also benchmarked against classical simulation methods, including tensor network (TN) approaches. One of the most prominent examples are matrix product states (MPS)~\cite{fannes1992,schollwock_review_2011}. While MPS are extremely powerful for one-dimensional (1D) systems, their application to 2D is typically limited to relatively small system sizes. This limitation does not apply to projected entangled-pair states (PEPS)~\cite{nishino01,gendiar03,verstraete2004,Murg_finitePEPS_07,Cirac_iPEPS_08,Xiang_SU_08,Verstraete_review_08,Orus_CTM_09,fu,Lubasch_conditioning,Orus_review_14,Corboz_varopt_16,Vanderstraeten_varopt_16,Fishman_FPCTM_17,Xie_PEPScontr_17,Corboz_Eextrap_16,Corboz_FCLS_18,Rader_FCLS_18,Rams_xiD_18,Hasik}, which provide a natural generalization of MPS to higher dimensions, including 3D~\cite{Vlaar3D,Charkiv3D}.

In Ref.~\onlinecite{Science_Dwave}, the coherent annealer was benchmarked against classical simulation methods in several QKZM scenarios for the random Ising model, demonstrating an advantage of the quantum device across a range of setups. In particular, for a cubic lattice, the main bottleneck of classical PEPS simulations was identified as the evaluation of expectation values.
In this work, we revisit this problem using both deterministic and Monte Carlo approaches. The latter allows us to estimate the final excitation energy of the system across the full range of quench times accessible to classical simulations. This development effectively shifts the computational bottleneck back to the time-evolution simulation itself, which must now be extended to cover a broader range of quench times.
Beyond the annealing scenario, the Monte Carlo scheme can also find applications in other problems, such as computing ground state properties~\cite{Vlaar3D,Charkiv3D}.

The remainder of the paper is organized as follows. In Sec.~\ref{sec:QA}, we briefly outline quantum annealing as implemented in the D-Wave quantum annealer. In Sec.~\ref{sec:ntu}, we describe the simulation of time evolution using a 3D tensor network. In Sec.~\ref{sec:DEV}, expectation values are computed deterministically within the network, while Sec.~\ref{sec:MC} presents an alternative approach based on Monte Carlo sampling. Finally, we conclude in Sec.~\ref{sec:concl}.

\begin{figure*}[t!]
\includegraphics[width=1.8\columnwidth]{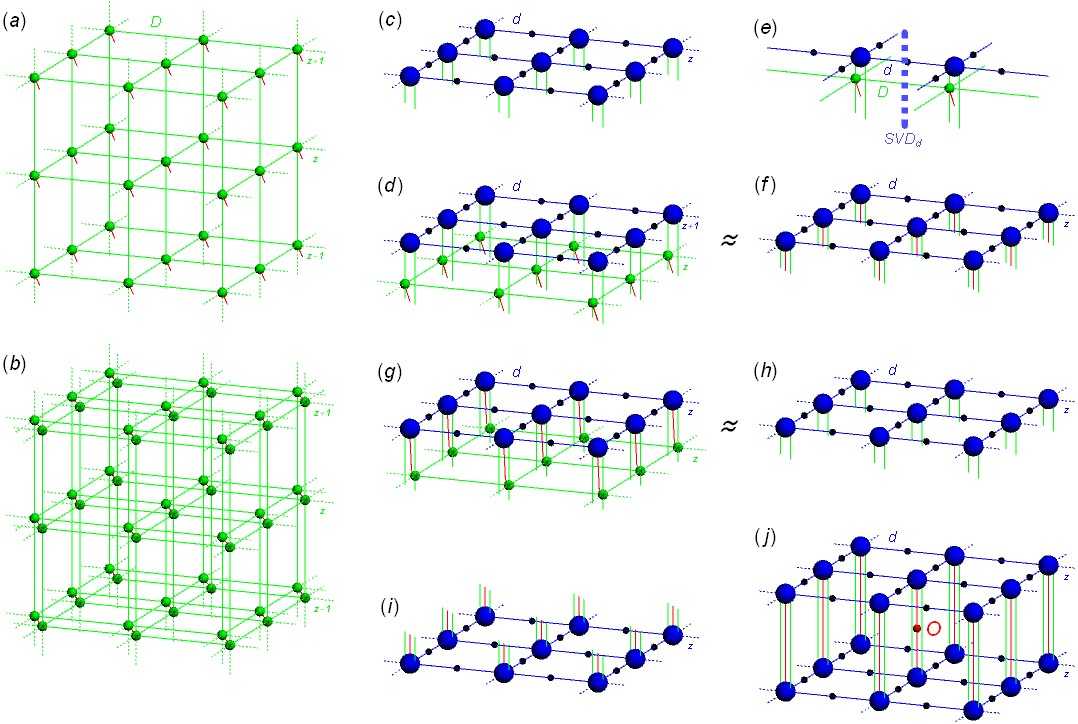}
\caption{
{\bf Deterministic expectation values in iPEPS. }
(a) A 3D infinite PEPS (iPEPS) tensor network with bond dimension $D$. The red lines denote physical indices.
(b) The norm squared of the state in (a), obtained by contracting the TN with its complex conjugate (indicated by stars) over the physical indices. Each horizontal layer at position $z$ can be interpreted as a transfer matrix ${\cal T}_z$.
(c) The upper boundary at layer $z$. The black dots denote diagonal matrices of singular values. The $z$-th boundary is obtained from the $(z+1)$-th boundary by applying the $z$-th transfer matrix ${\cal T}_z$, as illustrated in (d-g).
(d) The $z$-th iPEPS layer is applied to the $(z+1)$-th boundary, resulting in the $z$-th mixed boundary shown in (f), which carries additional physical indices.
(e) The bond dimension of the mixed boundary is truncated back to $d$ using the simple update, by truncating the singular values on each bond.
(g) The conjugate $z$-th iPEPS layer is applied to the mixed boundary (f), yielding the upper boundary at layer $z$. The bond dimension is again truncated to $d$ in (h).
(i) The lower mixed boundary at layer $z$ is obtained from the $(z-1)$-th lower boundary by applying the conjugate $z$-th iPEPS layer.
(j) Finally, the lower mixed boundary in (i) is contracted with the upper mixed boundary in (f) to make a double layer 2D network. An operator of interest, $O$, is inserted between physical indices at a selected site. The 2D network is contracted approximately with CTMRG as in Ref.~\cite{Corboz_tJ_CTMRG}.
}\label{fig:ipeps}
\end{figure*}

\section{Quantum annealing}
\label{sec:QA}

We consider a 3D transverse-field quantum Ising model on a cubic lattice. In a parametrization suitable for quantum annealing, its Hamiltonian is given by\cite{Science_Dwave}
\bea 
{\cal H}(s) &=& \Gamma(s)\;{\cal H}_D + {\cal J}(s)\; {\cal H}_I, 
\label{eq:H}\\ 
{\cal H}_D &=& -\sum_i \sigma^x_i,\\ {\cal H}_I &=& \sum_{ij} J_{ij} \sigma^z_i \sigma^z_j. 
\eea
Here, $\sigma^x_i$ and $\sigma^z_i$ denote Pauli operators acting on spin (qubit) $i$, and $s\in[0,1]$ is a dimensionless time-like parameter, see Fig. \ref{fig:QA}. In this work, we consider a cubic lattice with only nearest-neighbor (NN) couplings $J_{ij}$, randomly and uniformly distributed between $\pm 1$. The transverse field $\Gamma(s)$ and the energy scale of the Ising Hamiltonian ${\cal J}(s)$ are shown in Fig. \ref{fig:QA}. The parameter $s$ is ramped from $0$ to $1$ over the annealing time $t_a$, according to $s=t/t_a$. During this process, the Hamiltonian evolves from a nearly pure transverse-field Hamiltonian to the pure Ising Hamiltonian.

The evolution begins in the ground state of the initial Hamiltonian. Upon crossing the quantum phase transition from the paramagnetic phase to the spin-glass phase, excitations are generated in accordance with QKZM. The relevant critical exponents are $1/\nu = 1.55$ and $z = 1.3$ \cite{3D_KZ_DWave}. The residual excitation energy per spin in the Ising Hamiltonian at the end of the ramp is predicted \cite{3D_KZ_DWave} to scale as
\be
Q \propto t_a^{-(d\nu + z\nu - 1)/(1 + z\nu)} = t_a^{-0.965},
\label{eq:Q_KZ}
\ee
where $d = 3$ is the spatial dimensionality. 
The power law holds for long enough annealing time provided that the Kibble--Zurek correlation length imprinted on the state after crossing the critical point,
\be
\hat\xi \propto t_a^{0.35},
\label{eq:hatxiglass}
\ee
remains small compared with the system size.
In the following we attempt to test the energy power law with the 3D PEPS.

\begin{figure}[t!]
\includegraphics[width=0.99\columnwidth]{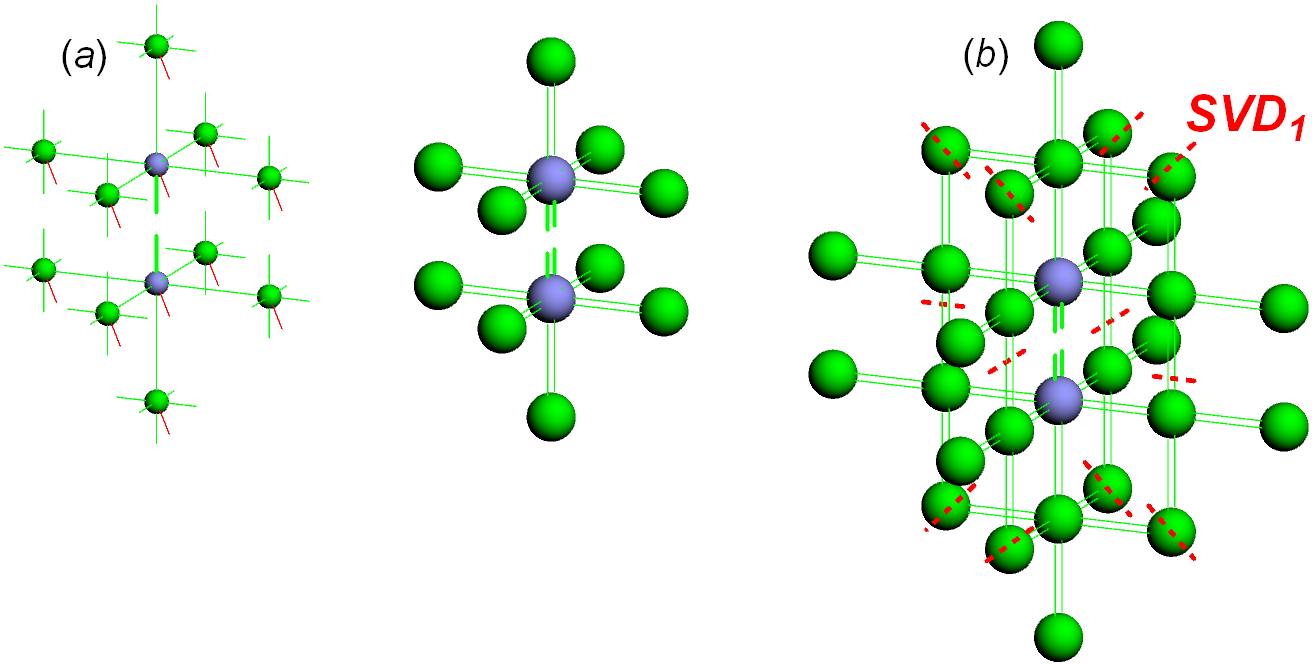}
\caption{
{\bf Trotter gate. }
(a, left) A fragment of the iPEPS shown in Fig. \ref{fig:ipeps}(a). The bluish tensors in the center have been acted upon by a two-site Trotter gate, which doubles the dimension of their shared bond, as indicated by the thicker lines. In this diagram, the doubled bond indices are left uncontracted.
(a, right) The Gram-Schmidt metric corresponding to the doubled indices, obtained as the overlap between the diagram on the left and its complex conjugate. The large bullets represent double PEPS tensors, each formed by contracting a PEPS tensor with its complex conjugate over the physical index and a subset of the corresponding bond indices. This metric is used to compress the doubled bond dimension back to $D$ as in Ref.~\onlinecite{ntu}.
(b) A larger fragment of the network and the corresponding metric tensor. To contract the metric efficiently, some double PEPS tensors and bonds are truncated via singular value decomposition, retaining only the leading singular value ($SVD_1$). This method is used in the present paper.
}\label{fig:ntu}
\end{figure}

\section{Time evolution}
\label{sec:ntu}

Figure~\ref{fig:ipeps}(a) shows the PEPS tensor-network ansatz on a cubic lattice. The diagram represents the probability amplitude $\psi_{i_1i_2...}$ of the state
\be
\sum_{i_1i_2...} \psi_{i_1i_2...} \ket{i_1i_2...}.
\ee
The amplitude is obtained by contracting the tensor network:
\be
\psi_{i_1i_2...} = {\rm Tr~} A^{i_1}_1 A^{i_2}_2 \dots.
\ee
Here $A^{i_j}_j$ is the tensor at site $j$ with physical index $i_j$, and Tr denotes contraction of the tensors over their virtual indices, as illustrated in Fig.~\ref{fig:ipeps}(a).
In two dimensions, iPEPS has been used to simulate sudden Hamiltonian quenches~\cite{CzarnikDziarmagaCorboz,HubigCirac,tJholeHubig,SUlocalization,SUtimecrystal,ntu,mbl_ntu,BH2Dcorrelationspreading,ising2D_correlationsperading,Mazur_BH,Corboz_SF}.
The neighborhood tensor update (NTU)~\cite{ntu}, that is employed here, has been applied to simulations of many-body localization~\cite{mbl_ntu}, Kibble-Zurek ramps~\cite{schmitt2021quantum,Mazur_BH,Science_Dwave,XX_KZ}, the bang-bang preparation of ground states using shallow quantum circuits~\cite{BB_iPEPS,BB2}, as well as thermal states obtained by imaginary-time evolution in the fermionic Hubbard model~\cite{Hubbard_Sinha,Sinha_Wietek_Hubbard,tJ_Zhang}.

Unitary time evolution of the 3D PEPS is implemented using $n$ finite time steps of equal duration $dt$,
\be
e^{-idt {\cal H}(s_1)}
\dots
e^{-idt {\cal H}(s_n)},
\ee
where $t_a=n~dt$ is the total anneal time and $s_k=(k-1/2)dt/t_a$. Within the second-order Suzuki-Trotter decomposition, we approximate
\be
e^{-idt {\cal H}(s_k)}
\approx
e^{-\frac{i dt}{2}\Gamma(s_k) {\cal H}_D}
e^{- i dt{\cal J}(s_k) {\cal H}_I}
e^{-\frac{i dt}{2}\Gamma(s_k) {\cal H}_D} .
\ee
Here the driving term is a product of one-site gates:
\be
e^{-\frac{i dt}{2}\Gamma(s_k){\cal H}_D} =
\prod_j e^{\frac{i dt}{2}\Gamma(s_k)\sigma^x_j}.
\ee
Its application to the ansatz in Figure~\ref{fig:ipeps}(a) is straightforward: the (transpose of each) one-site gate is applied to the corresponding physical index of the tensor network. The interaction term is a product of mutually commuting NN Ising gates:
\bea
e^{-i dt{\cal J}(s_k) {\cal H}_I} =
\prod_{\langle ij\rangle}
e^{-i dt{\cal J}(s_k) J_{ij} \sigma^z_i\sigma^z_j }.
\eea
Each NN Trotter gate can be written as a sum of two terms:
\be
e^{-i dt{\cal J}(s_k) J_{ij} \sigma^z_i\sigma^z_j } =
1_i 1_j \cos \phi_{ij} 
- i \sigma^z_i\sigma^z_j \sin \phi_{ij} ,
\ee
where $\phi_{ij}=dt{\cal J}(s_k) J_{ij}$. This sum can be expressed in the more compact form
\be
\sum_{m=1,2} T_i^m T_j^m,
\ee
where the one-site operators are $ T_i^1 = 1_i \sqrt{\cos\phi_{ij}} $ and $ T_i^2 = \sigma^z_i \sqrt{-i\sin\phi_{ij}} $. The resulting expression can be viewed as a contraction of two tensors over a common bond index $m$ of dimension $2$.
Therefore, an Ising Trotter gate applied to a nearest-neighbor (NN) bond doubles the bond dimension from $D$ to $2D$. To prevent the exponential growth of the bond dimension during time evolution, it must be truncated back to $D$. This truncation is performed using the neighborhood tensor update (NTU), for numerical details see Ref.~\onlinecite{ntu}.

Figure \ref{fig:ntu} illustrates two of the tensor neighborhoods used in the NTU truncation \cite{Science_Dwave}. On the one hand, larger neighborhoods provide more accurate truncations, as the resulting Gram-Schmidt metric tensor more faithfully represents the tensor environment, enabling more efficient use of the limited bond dimension. On the other hand, the computational cost of exact tensor contraction increases rapidly with neighborhood size. Unlike in two dimensions \cite{ntu}, in 3D we cannot afford neighborhoods containing closed loops. These loops are approximated using SVD$_1$ truncations. 
This mean-field-like approximation is expected to provide a better description than neglecting loops altogether, as in the simple-update scheme~\cite{Orus_review_14} or its belief-propagation variant~\cite{Tindall_BP}.
With or without the SVD$_1$ approximations, the resulting metric tensor remains Hermitian and non-negative, ensuring stable bond-dimension truncation. 
The contraction of both Gram-Schmidt metric tensors presented in Fig. \ref{fig:ntu} has a computational cost scaling as ${\cal O}\left(D^8\right)$.
In this paper we employ the method in Fig. \ref{fig:ntu}(b).

The time evolution is performed for a range of ramp times for both an infinite lattice with periodic randomness and an $L^3$ unit cell, and a finite lattice with open boundary conditions (PBC and OBC). The time step is $dt=0.01\mathrm{ns}$, which is sufficiently small for the two coefficients shown in Fig.~\ref{fig:QA}, particularly in the regime where their magnitudes are comparable. The residual excitation energy near the end of the ramp - specifically at $s = 0.6$, when the transverse field is already negligible - is evaluated using deterministic methods for PBC~\cite{Vlaar3D,Charkiv3D} and Monte Carlo methods for OBC. It is defined as
\be
Q=
{\cal N}^{-1}
\sum_{\langle ij\rangle}
\left[
\left\langle\sigma^z_i\sigma^z_j\right\rangle-
\left\langle\sigma^z_i\sigma^z_j\right\rangle_{\rm GS}
\right].
\label{eq:Q}
\ee
Here, the sum runs over all NN bonds-within the unit cell for PBC and over the open lattice for OBC-and ${\cal N}$ is the total number of bonds. $\langle\dots\rangle$ denotes the expectation value in the quantum state near the end of the ramp, while $\langle\dots\rangle_{\rm GS}$ denotes the value in the ground state of the classical Hamiltonian ${\cal H}_I$ in Eq.~\eqref{eq:H}.
Anticipating the KZ correlation length \eqref{eq:hatxiglass} to remain much shorter than $L$ even for the longest considered anneal times, the number of disorder samples is one. The results are effectively self-averaged over the disorder within the $L\times L\times L$ lattice (or unit cell). The size $L=8$ is considered. 

\begin{figure}[t!]
\includegraphics[width=0.9\columnwidth]{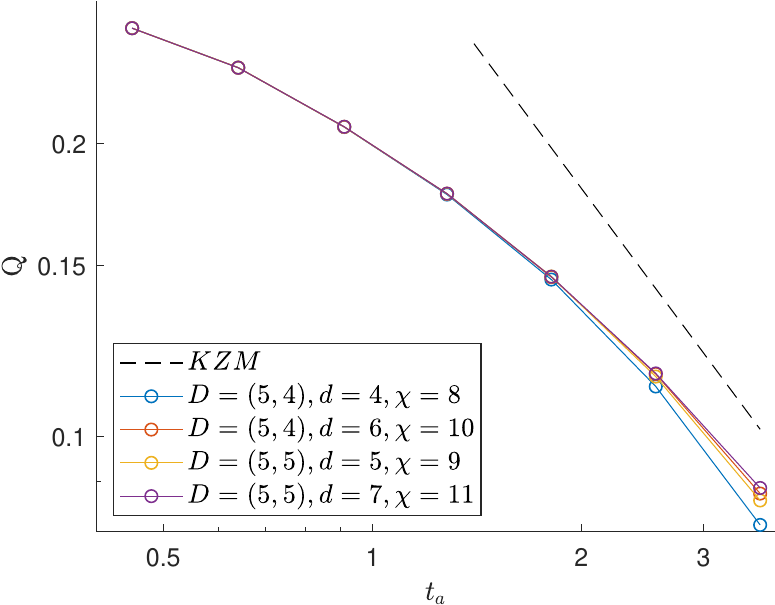}
\caption{
{\bf Excitation energy - deterministic. }
Average residual excitation energy per bond, $Q$ (in GHz), versus annealing time $t_a$ (in ns). The dashed line shows the slope predicted by the Kibble-Zurek power law \eqref{eq:Q_KZ}, expected to hold for sufficiently long annealing times, provided that the KZ correlation length \eqref{eq:hatxiglass} remains much smaller than $L$. The energies were computed using the deterministic method for an infinite lattice with a periodic unit cell of size $L^3=8^3$.
}\label{fig:det_E}
\end{figure}

\begin{figure}[t!]
\includegraphics[width=0.9\columnwidth]{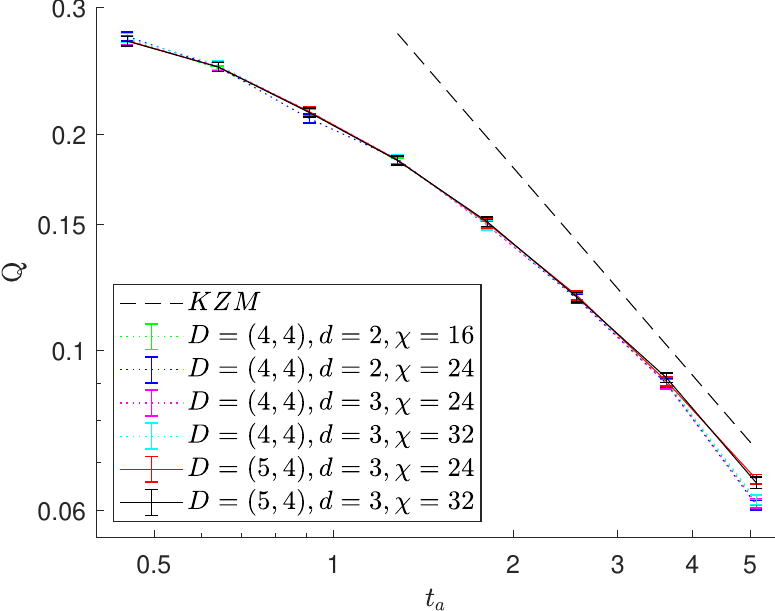}
\caption{
{\bf Excitation energy - Monte Carlo. }
Average residual excitation energy per bond, $Q$ (in GHz), as a function of the annealing time (in ns) for an open cubic lattice of linear size $L=8$. The dashed line indicates the predicted Kibble-Zurek power-law scaling, expected to hold for sufficiently long annealing times, provided that the KZ correlation length \eqref{eq:hatxiglass} remains much smaller than $L$.
}\label{fig:MC_E}
\end{figure}

The tensor network is evolved with a fixed bond dimension $D$ throughout the dynamics. In the final state, it is occasionally truncated to a smaller bond dimension to facilitate the evaluation of the energy. The final truncation is done in the same way as after a Trotter gate. In Figs. \ref{fig:det_E} and \ref{fig:MC_E}, this is denoted as $D=(D_e,D_t)$, where $D_e$ is the bond dimension used during the evolution and $D_t$ is the truncated bond dimension used for the final energy evaluation.

\section{Deterministic expectation values}
\label{sec:DEV}

The deterministic evaluation combines the methods of Ref.~\onlinecite{Vlaar3D} and Ref.~\onlinecite{Charkiv3D}, as illustrated in Fig.~\ref{fig:det_E}. Panel (a) shows the iPEPS ansatz with a periodic unit cell of size $L^3$ corresponding to the periodicity of the disorder. In (b), the iPEPS is contracted through the physical indices with its conjugate to form a tensor network representing its norm squared. The $z$-th horizontal layer of the norm is treated as a transfer matrix ${\cal T}_z$. The transfer matrices are periodic, ${\cal T}_{z\pm L}={\cal T}_z$, and each transfer matrix has a periodic $L^2$ unit cell. We seek the leading upper and lower eigenvectors, $\left|{\cal U}_z\right)$ and $\left({\cal L}_z\right|$, represented by the iPEPS shown in panels (c) and (i), respectively. For the leading eigenvalue $\Lambda$, they satisfy
\bea
\Lambda
\left|{\cal U}_z\right) &=&
{\cal T}_z\dots{\cal T}_{z+L-1}
\left|{\cal U}_z\right), \label{eq:calU}\\
\Lambda
\left({\cal L}_z\right| &=&
\left({\cal L}_z\right|
{\cal T}_{z-L+1}\dots{\cal T}_{z}. \label{eq:calL} 
\eea
These boundary iPEPS have a periodic $L^2$ unit cell and bond dimension $d$. Along the vertical direction, with layers labeled by the index $z$, there is a periodic sequence of $L$ distinct upper and lower boundary states. These boundary states are obtained using the power method suggested by \eqref{eq:calU} and \eqref{eq:calL}. Each horizontal layer of the double-PEPS network in Fig.~\ref{fig:ipeps}(b) acts as a transfer matrix that is repeatedly applied to the boundary state until convergence is achieved. Since ${\cal T}_z$ is double-layered, its application to ${\cal U}_{z+1}$ is carried out in two steps, as shown in Fig.~\ref{fig:ipeps}(d,f,g,h). First, only the $z$-th iPEPS layer is applied, as shown in Fig.~\ref{fig:ipeps}(d). After this application, the boundary bond dimension is truncated back to $d$ in Fig.~\ref{fig:ipeps}(f) to prevent uncontrolled growth. The resulting truncated iPEPS is a mixed upper boundary, 
$\left|U_z\right)$, with additional physical indices. Next, the $z$-th conjugate iPEPS layer is applied, as shown in Fig.~\ref{fig:ipeps}(g). After a second truncation, it becomes the upper boundary $\left|{\cal U}_z\right)$ shown in Fig.~\ref{fig:ipeps}(h).

The truncations are performed using the simple-update (SU) scheme, as illustrated in Fig.~\ref{fig:ipeps}(e). For numerical details, see Ref.~\cite{orus_review_2014} for a general discussion and Refs.~\cite{Vlaar3D,Charkiv3D} for the present context.
To perform the SVD efficiently, the left and right halves of the diagram are first decomposed using QR factorization, yielding reduced matrices whose product is then subjected to an SVD. The computational cost of the QR decomposition is ${\cal O}\left(D^7 d^5 p\right)$, where $p=2$ is the physical dimension.

After all $L$ upper and lower boundaries have converged, the corresponding mixed boundaries are contracted together as $\left(L_z\right|\left.U_z\right)$, see Fig.~\ref{fig:ipeps}(j).
The result is a two-dimensional tensor network with an $L\times L$ unit cell. It is approximately contracted using the corner transfer matrix renormalization group (CTMRG) \cite{Orus_CTM_09,Corboz_tJ_CTMRG,Corboz_varopt_16} method with bond dimension $\chi$. The CTMRG procedure for a periodic network with an $L^2$ unit cell is described in the Appendix of Ref.~\onlinecite{Corboz_tJ_CTMRG}. The leading computational cost of CTMRG arises from singular-value decompositions of enlarged corner tensors and scales as ${\cal O}\left(d^6\chi^3\right)$.

The residual energy as a function of the annealing time is shown in Fig.~\ref{fig:ipeps} on a log-log scale. For annealing times approaching $3\mathrm{ns}$, the data begin to approach the slope predicted by the Kibble-Zurek power law \eqref{eq:Q_KZ}, which is expected to hold for sufficiently long annealing times, provided that the KZ correlation length \eqref{eq:hatxiglass} remains much smaller than $L$.

The evaluation of the upper and lower boundary states constitutes the primary computational bottleneck of the method, as it relies on large singular value decompositions. In the present calculations, the largest bond dimension $d=7$ was chosen such that the truncation error - defined as the sum of discarded singular values divided by the sum of all singular values in Fig.~\ref{fig:ipeps}(e) - remains below $0.01$, even for the slowest annealing ramps. For each pair $D,d$, the energy is converged with respect to the CTMRG bond dimension $\chi$.

\begin{figure}[t!]
\includegraphics[width=0.7\columnwidth]{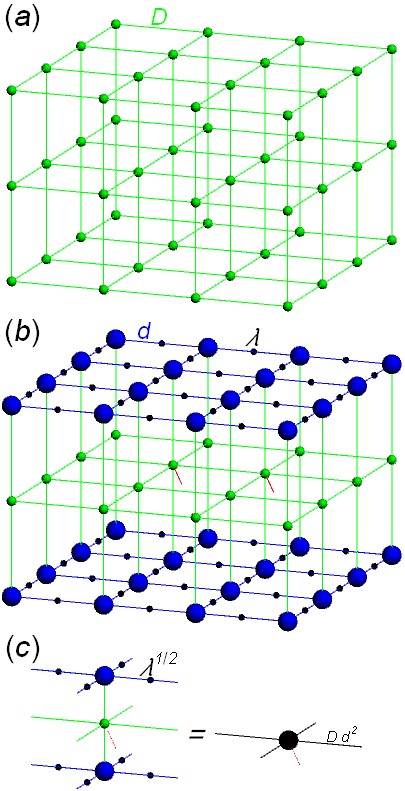}
\caption{
{\bf Monte Carlo sampling in a 3D PEPS. }
(a) A projected PEPS in which all physical indices have been assigned definite values. Unlike the full PEPS shown in Fig.~\ref{fig:ipeps}(a), the fixed physical indices are not displayed. Here, a $3\times3\times4$ PEPS is shown for illustration.
(b) The upper and lower projected PEPS boundaries (blue) are used to compute the probability amplitudes for two selected sites with the full PEPS tensors. These amplitudes depend explicitly on the physical indices of the sites (red).
(c) At each site, a triple tensor is defined, which may include or exclude the physical index depending on whether the full or projected PEPS tensor is considered.
}\label{fig:3d_peps}
\end{figure}

\begin{figure}[t!]
\includegraphics[width=0.7\columnwidth]{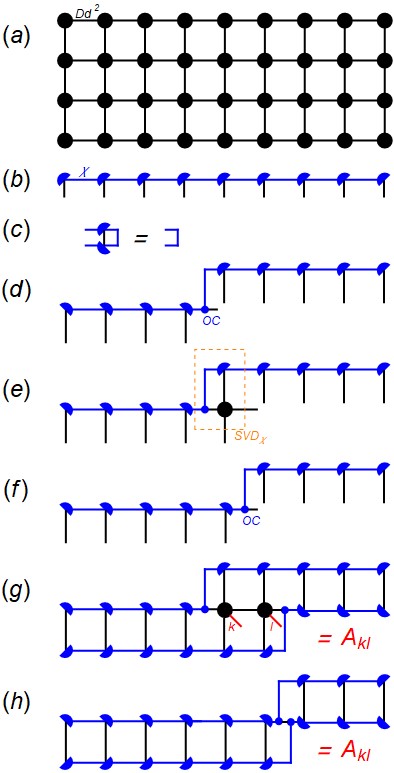}
\caption{
{\bf Monte Carlo sampling in a 2D layer. }
(a) The planar network corresponding to the triple-layer network in Fig. \ref{fig:3d_peps}(b), expressed using the tensors defined in Fig. \ref{fig:3d_peps}(c). Each row of the network can be interpreted as a transfer matrix.
(b) An upper boundary of the network.
(c) The boundary in right-canonical form, with isometries satisfying the orthogonality condition shown.
A new upper boundary is obtained by applying a single row transfer matrix. Instead of applying the entire transfer matrix at once and then compressing the bond dimension, the application proceeds via a zipper procedure \cite{metts_peps}, where the transfer matrix tensors are applied one by one, as illustrated in (d-f).
(d) A mixed boundary halfway through the process. It is left-right canonical, with the tensor in the middle acting as the orthogonality center (OC).
(e) The singular value decomposition (SVD) of the orange rectangle, truncated to the $\chi$ leading singular values, moves the OC one site to the right.
(f) The truncated SVD interpolates between two mixed boundaries, resulting in a new intermediate boundary.
(g) A mixed upper boundary and a mixed lower boundary are used to compute probability amplitudes $A_{kl}$ for different combinations of physical indices $k$ and $l$.
(h) For each $k,l$, the calculation is performed by moving the upper OC two sites to the right and contracting the resulting diagram.
After updating a pair of indices, the upper and lower OCs are shifted one site to the right to calculate the amplitudes for the next overlapping pair. 
Repeating this procedure row by row and layer by layer completes a full Monte Carlo sweep of the 3D lattice.  
}\label{fig:sampling}
\end{figure}

\begin{figure}[t!]
\includegraphics[width=0.7\columnwidth]{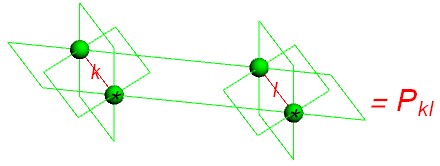}
\caption{
{\bf Local proposal function. }
Candidate values of the physical index pair $(k,l)$ are proposed using a proposal function derived from the local probability $P_{k,l}$ shown in the figure. In this diagram, the indices $(k,l)$ are fixed, and no summation over them is performed.
}\label{fig:trial}
\end{figure}

\section{Monte Carlo sampling}
\label{sec:MC}

We adapt the Monte Carlo (MC) Metropolis-Hastings algorithm to evaluate expectation values on a finite $L^3=8^3$ lattice with open boundary conditions. Its main advantage is that, instead of using the full PEPS shown in Fig.~\ref{fig:ipeps}(a), the algorithm operates on a projected PEPS in which the physical indices are fixed, as illustrated in Fig.~\ref{fig:3d_peps}(a). This eliminates the need for a double-layer network, such as the one shown in Fig.~\ref{fig:ipeps}(b).

The projected PEPS in Fig.~\ref{fig:3d_peps}(a) represents the probability amplitude for a given configuration of physical indices. Each horizontal layer can be regarded as a transfer matrix ${\cal T}_z$, with $z=1,\ldots,L$. To treat all layers on an equal footing, we assume that, in the special case of $z=1$, ${\cal T}_1$ has upper indices of dimension $D$ but is nonzero only when these upper indices take the value $1$. Similarly, ${\cal T}_L$ has lower indices of dimension $D$. We can then define upper and lower PEPS boundary states, $\left|{\cal U}_z\right)$ and $\left({\cal L}_z\right|$, for $z=0,\ldots,L+1$. Each is represented by an $L\times L$ PEPS with bond dimension $d$ and ``physical'' dimension $D$. The states $\left|{\cal U}_0\right)$ and $\left({\cal L}_{L+1}\right|$ are trivial, i.e., their tensors are nonzero only when all bond indices and the physical index are equal to $1$. Starting from the trivial $\left|{\cal U}_0\right)$ (or $\left({\cal L}_{L+1}\right|$), the upper (lower) boundary states are obtained by successive application of the transfer matrices, with each application followed by SU truncation of the bond indices:
\bea
\left|{\cal U}_z\right)
& \stackrel{\rm SU}{\approx} &
{\cal T}_z \left|{\cal U}_{z-1}\right), \label{eq:Uz}\\
\left({\cal L}_z\right|
& \stackrel{\rm SU}{\approx} &
\left({\cal L}_{z+1}\right|
{\cal T}_z.                            \label{eq:Lz}
\eea
Importantly, the bond dimension $d$ of these boundary states---whose computation remains the main computational bottleneck---can be chosen to be smaller than that required for the double-layer network shown in Fig.~\ref{fig:ipeps}(c).
With the upper and lower boundaries at hand, the probability amplitude in Fig.~\ref{fig:3d_peps}(a) can be evaluated as
\be
\left({\cal L}_{z+1}\right| {\cal T}_z \left|{\cal U}_{z-1}\right),
\ee
i.e., by sandwiching a layer of the projected PEPS between the corresponding upper and lower boundaries, as shown in Fig.~\ref{fig:3d_peps}(b).
The SU updates in Eqs.~\eqref{eq:Uz} and \eqref{eq:Lz} are performed in a manner similar to that shown in Fig.~\ref{fig:ipeps}(c--h), although there is no need to introduce mixed boundaries. The computational cost of the QR decomposition in the simple update is ${\cal O}\left(D^6d^5\right)$, i.e., formally lower by a factor of $Dp$ than in the double-layer deterministic method. Furthermore, the smaller required bond dimension $d$ yields an additional reduction in the computational cost.

The trade-off is the need for sampling to estimate observables. Here, we sample in the computational $\sigma^z$ eigenbasis, though other bases can be accessed via a suitable unitary transformation of the physical indices. Sampling proceeds by updating pairs of physical indices. Fig. \ref{fig:3d_peps}(b) shows a triple-layer network representing amplitudes for different combinations of a chosen pair of indices. This triple-layer network is converted into a planar network (Fig. \ref{fig:sampling}(a)) by defining planar tensors as in Fig. \ref{fig:3d_peps}(c).

The $L\times L$ planar network is approximately contracted by updating its upper and lower matrix product state (MPS) boundaries, as outlined in Figs.~\ref{fig:sampling}(d--f), using the zipper method described in detail in Ref.~\cite{metts_peps}. Each row $x$ of the planar network in Fig.~\ref{fig:sampling}(a) is treated as a transfer matrix ${\cal T}_x$. The transfer matrices are repeatedly applied to the upper and lower boundaries, $\left|{\cal U}_x\right)$ and $\left({\cal L}_x\right|$, with $x=0,\dots,L+1$. An upper boundary is shown in Fig.~\ref{fig:sampling}(b). It is an MPS in right-canonical form, with isometries satisfying the right-canonical gauge condition shown in Fig.~\ref{fig:sampling}(c). For an introduction to matrix product states, see the review in Ref.~\cite{Vanderstraeten2019}. In general, each new boundary is obtained from the previous one by applying the transfer matrix, followed by a truncation of the bond dimension:
\bea
&&
{\cal T}_x \left|{\cal U}_{x-1}\right)
\stackrel{\rm zip.}{\approx}
\left|{\cal U}_{x}\right),~~~~
\left({\cal L}_{x+1}\right| {\cal T}_x
\stackrel{\rm zip.}{\approx}
\left({\cal L}_{x}\right|.
\eea
However, in the zipper method the matrix product operator (MPO) ${\cal T}_x$ is not applied in its entirety and truncated only afterward, bond by bond. Instead, during a left-to-right sweep, one MPO tensor is applied at a time, as shown in Fig.~\ref{fig:sampling}(e), followed immediately by SVD truncation.
The SVD constitutes the leading computational cost of the zipper method, scaling as ${\cal O}\left(D^3d^6\chi^3\right)$.
A by-product of this procedure is the mixed boundary obtained halfway through each transfer-matrix application, as shown in Fig.~\ref{fig:sampling}(d,f). These mixed boundaries are in the mixed-canonical form, with the tensors to the right (left) of the orthogonality center (OC) in the right- (left-) canonical form. By repeatedly applying the transfer matrix to the upper and lower boundaries, one obtains $L$ upper and lower boundaries together with the corresponding intermediate mixed boundaries. The upper and lower mixed boundaries can then be combined to evaluate probability amplitudes for different spin configurations. Figure~\ref{fig:sampling}(g) shows the diagram for a two-spin probability amplitude $A_{kl}$, assuming that all other spins are fixed. Its value is obtained by shifting the upper orthogonality center (OC) in Fig.~\ref{fig:sampling}(g) two sites to the right and then contracting the upper and lower boundaries, as shown in Fig.~\ref{fig:sampling}(h). This amplitude is then used to update the two spins.

As in the Metropolis-Hastings MC algorithm, new values of the indices $(k,l)$ are proposed according to a simple proposal function based on the local probability $\propto P_{kl}$ (Fig.~\ref{fig:trial}). The target probability, $\propto |A_{kl}|^2$, then determines whether the proposed indices are accepted.
We do not employ the more accurate proposal functions considered in 2D~\cite{MC_PEPS}, as their evaluation incurs substantial computational overhead, while even this simplest proposal function already yields short autocorrelation times.

\begin{figure}[t!]
\includegraphics[width=0.9\columnwidth]{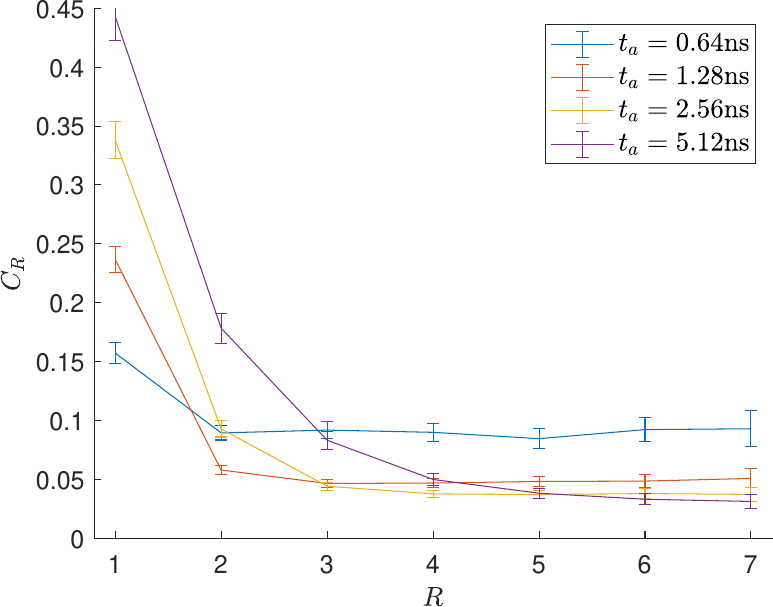}
\caption{
{\bf Correlation function - Monte Carlo. }
The correlator \eqref{eq:CR} for several values of the annealing time. Its range is increasing with the time, compare the Kibble-Zurek prediction \eqref{eq:hatxiglass}, but remains smaller than $L=8$.
}\label{fig:MC_CR}
\end{figure}

The residual energy as a function of the annealing time is shown in Fig. \ref{fig:MC_E} on a log-log scale. For annealing times around $5$ns, the data approach the slope predicted by the Kibble-Zurek power law \eqref{eq:Q_KZ}. This residual energy decreases for longer annealing times and therefore requires more extensive sampling to achieve the same relative accuracy. The MC sweeps are performed layer by layer and, within each layer, row by row. There are three possible layer orientations and, for each of them, two row orientations. Equivalently, there are six different ways to sweep the 3D lattice, which are applied sequentially. The number of MC sweeps increases from $16\times6$ for the fastest ramps to $420\times6$ for the slowest ones in order to maintain the same relative error estimate $\delta Q/Q=0.01$. The autocorrelation time is on the order of a few sweeps. To estimate the error bars, the data are averaged over blocks of 6 sweeps, and the statistical error is then calculated from these block averages.

\begin{figure}[t!]
\includegraphics[width=0.9\columnwidth]{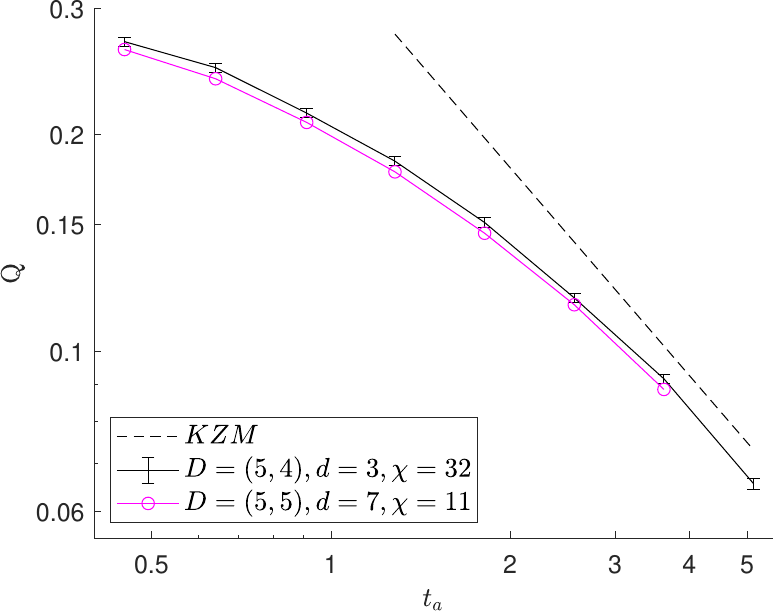}
\caption{
{\bf Excitation energy - comparison. }
Residual excitation energy per bond, $Q$ (in GHz), as a function of the annealing time (in ns) for the periodic/open lattice evaluated by deterministic/MC methods. Only the results from Figs.~\ref{fig:det_E} and \ref{fig:MC_E} corresponding to the largest bond dimensions are shown.
}\label{fig:MC_E_comparison}
\end{figure}

Having accumulated all the MC samples, it is straightforward to estimate the correlation function
\be
C_R=
{\cal N}^{-1}_R
\sum_{\langle i,j \rangle_R}
\left|
\langle \sigma^z_{i} \sigma^z_{j} \rangle
\right|.
\label{eq:CR}
\ee
Here, the sum runs over all pairs of sites $i,j$ separated by a distance $R$ along $x,y$ or $z$ direction, and ${\cal N}_R$ is the number of such pairs. Fig. \ref{fig:MC_CR} shows the correlator obtained from the MC-estimated expectation values. According to the KZ mechanism, its range should increase as \eqref{eq:hatxiglass}. The correlators for $t_a=2.56,5.12$ns are not inconsistent with this prediction. Importantly, for self-consistency, these correlators are still short-ranged when compared to the size of the lattice.

Finally, in Fig. \ref{fig:MC_E_comparison}, we compare the most accurate results from Figs. \ref{fig:det_E} and \ref{fig:MC_E}. Although there is a slight systematic difference between the infinite lattice with periodic disorder and the open lattice, both datasets exhibit the same dependence. For the deterministic evaluation, the missing $5.12\mathrm{ns}$ data point proved to be particularly time-consuming because its long correlations led to slow convergence of the computationally expensive upper and lower boundaries. In contrast, the Monte Carlo data extend into the Kibble-Zurek scaling regime.

\section{Conclusion}
\label{sec:concl}
In Ref. \cite{Science_Dwave}, the estimation of expectation values in the 3D tensor network was identified as the main bottleneck in tensor-network simulations of quantum annealing. In this work, we show that observables can be evaluated either using deterministic methods or, more efficiently, via 3D Monte Carlo sampling. The sampling over an open-boundary network proceeds layer by layer and, within each layer, row by row, employing 2D PEPS and 1D MPS boundaries, respectively. As a result, the computational bottleneck shifts to the simulation of the time evolution. This bottleneck is less restrictive in the case of imaginary-time evolution toward the ground state \cite{Vlaar3D,Charkiv3D}, which is where the Monte Carlo approach can be particularly advantageous.

The advantage of the Monte Carlo approach stems from the single-layer structure of the involved tensor networks, which drastically reduces the contraction cost compared to the double-layer networks required in deterministic evaluations. In the application considered here, the computational gain from the single-layer structure outweighs the cost of multiple Monte Carlo sweeps required to achieve the desired error bars.

The data used for Figs. \ref{fig:det_E} and \ref{fig:MC_E} are openly available from https://doi.org/10.57903/UJ/XQWNGX.

\begin{acknowledgments}
I am indebted to Marek Rams and Yintai Zhang for stimulating discussions.
This research was funded by the National Science Centre (NCN), Poland, under projects 2024/55/B/ST3/00626.
The research was also supported by a grant from the Priority Research Area DigiWorld under the Strategic Programme Excellence Initiative at Jagiellonian University (MMR,JD).
\end{acknowledgments}

\bibliographystyle{apsrev4-2}
\bibliography{KZref.bib}


\end{document}